\documentstyle[11pt,psfig,aaspp4]{article}
\def\be{\begin{equation}}
\def\ee{\end{equation}}
\def\etal{{\it et al. }}
\def\kms{km~s$^{-1}$~}

\def\nin+{584~}

\def\W50{W$_{50}$~}
\journalid{337}{Someday 1999}
\articleid{11}{14}

\slugcomment{To apper in {\it The Astrophysical Journal (Letters)}}

\begin{document}
\hskip 3.5in{Version 3.0 \hskip 10pt 27 Jul 1998}
\title{Peculiar Velocity Dipoles of Field Galaxies}

\author {Riccardo Giovanelli, Martha P. Haynes}
\affil{Center for Radiophysics and Space Research
and National Astronomy and Ionosphere Center\altaffilmark{1},
Cornell University, Ithaca, NY 14853}

\author {Wolfram Freudling}
\affil{Space Telescope--European Coordinating Facility and European Southern 
Observatory, Karl--Schwarzschild--Str. 2, D--85748 Garching b. M\"unchen, 
Germany}

\author {Luiz N. da Costa}
\affil{European Southern Observatory, Karl--Schwarzschild--Str. 2, D--85748
Garching b. M\"unchen, Germany and Observatorio Nacional, Rio de Janeiro,
Brazil}

\author {John J. Salzer}
\affil{Astronomy Dept., Wesleyan University, Middletown, CT 06459}

\author {Gary Wegner}
\affil{Dept. of Physics and Astronomy, Dartmouth College, Hanover, 
NH 03755}

\altaffiltext{1}{The National Astronomy and Ionosphere Center is
operated by Cornell University under a cooperative agreement with the
National Science Foundation.}

\hsize 6.5 truein
\begin{abstract}

The Tully--Fisher (Tully \& Fisher 1977; TF) relation is applied to obtain 
peculiar velocities of field spirals galaxies and to calculate dipoles 
of the peculiar velocity field to $cz \simeq 8000$ \kms. The field galaxy 
sample is spatially 
co--extensive with and completely independent on a cluster sample, for 
which dipole characteristics are given in a separate paper. Dipoles of 
the peculiar velocity field  are obtained separately by applying (i)
an inverse version of the TF relation and selecting galaxies by
redshift windowing and (ii) a direct TF relation, with velocities 
corrected for the inhomogeneous Malmquist bias, and windowing
galaxies by TF distance. The two determinations agree,
as they do with the cluster sample. When measured in a reference frame 
in which the Local Group is at rest, the dipole moment of field galaxies
farther than $\sim 4000$ \kms is in substantial agreement, both in amplitude 
and direction, with that exhibited by the Cosmic Microwave Background
radiation field.

\end{abstract}

\keywords{galaxies: distances and redshifts  -- cosmology: 
observations; cosmic microwave background; distance scale }

\section{Introduction}

It is generally assumed that the Doppler shift arising from the solar 
motion is responsible for the Cosmic Microwave Background (CMB) radiation 
dipole moment. Allowing for solar motion with 
respect to the Local Group (LG) of galaxies, the CMB dipole
(Lineweaver \etal 1996) translates into a velocity ${\bf V}_{cmb}$ of the 
LG with respect to the comoving reference frame, of amplitude $611\pm22$ 
\kms, directed towards $l=273^\circ\pm 3^\circ$, $b=27^\circ\pm 3^\circ$. 
Most of the uncertainty arises from that on the motion 
of the Sun with respect to the LG, which we assume to have an amplitude of 
300 \kms and directed towards $l=90^\circ$, $b=0^\circ$ (de Vaucouleurs, de 
Vaucouleurs \& Corwin 1976).

In linear theory, the peculiar velocity induced on the LG by the
inhomogeneities present within a sphere of radius $R$ is
\be
{\bf V}_{pec,LG}(R) = {H_\circ \Omega_\circ^{0.6}\over 4\pi} 
\int \delta_{mass}({\bf r})
{{\bf r}\over r^3} W(r,R) d^3{\bf r}     
\ee
where $W(r,R)$ is a window function of width $R$, $H_\circ {\bf r}$ is
the distance in \kms, $\delta_{mass}$ is the mass overdensity at {\bf r}
and $\Omega_\circ$ is the cosmological density parameter. Assuming that 
the CMB dipole is the result of a Doppler shift, then there must be 
identity between ${\bf V}_{cmb}$ and ${\bf V}_{pec,LG}(R)$ as 
$R\rightarrow \infty$. As $R$ increases, ${\bf V}_{pec,LG}(R)$ converges 
towards ${\bf V}_{cmb}$ if the average value of $\delta_{mass}$ within 
a shell of radius 
$R$ approaches zero. In a Universe which on large scales is homogeneous,
it is thus reasonable to expect that the {\it reflex} motion of the
LG, with respect to the contents of a shell of large enough radius $R$,
will exhibit a dipole that closely matches that of the CMB radiation
field. How large should $R$ be, for that convergence to be observed?
The issue is widely debated, and positions can roughly be divided ---
with much grey area in between --- between two main camps: one in which
the vast majority of the local dynamics is determined by mass fluctuations
within $cz \simeq$ 5--10,000 \kms, and one in which a very substantial
fraction of that motion arises outside $cz \simeq 10,000$ \kms. The latter
is substantiated by several studies, from the early suggestion of Scaramella 
\etal (1989) on the importance of the Shapley Supercluster to the results 
of Lauer \& Postman (1994), based on the reference frame defined by 119 
Abell clusters within roughly $cz \sim$ 15,000 \kms. Lauer \& Postman 
found the reflex motion of the LG with respect to the cluster sample to 
be given by a vector ${\bf V}_{lp}$ of amplitude $561\pm284$ \kms,
directed towards $(l,b)=(220^\circ,-28^\circ)\pm 27^\circ$. The lack of 
coincidence between ${\bf V}_{lp}$ and ${\bf V}_{cmb}$ implies an overall 
bulk flow of the Lauer \& Postman cluster reference frame of $689\pm178$ 
\kms towards $(l,b)=(343^\circ,+53^\circ)$. This result was confirmed by 
a re--analysis of the Lauer \& Postman data by Colless (1995), but was 
found in conflict with the studies of Riess \etal (1995) and Giovanelli 
\etal (1996). 

Here we analyze the dipole signatures of the peculiar velocity field of 
the SFI sample of field spirals. The characteristics of the sample are 
briefly described in Section 2; for further details, see Giovanelli \etal 
(1994) and the data presentation in Haynes \etal (1998a,b).  In Section 3 
we present the results of the dipole fits and discuss their amplitudes and 
significance in connection with the issue of the convergence depth of the 
local Universe.

\section{SFI Samples and Peculiar Velocity Calculations}   

The SFI sample was selected by adopting strict angular size limits for
the target galaxies, which varied with redshift in order not to underpopulate
more distant shells (see Giovanelli \etal 1994).
Galaxies observed by our team North of $\sim -35^\circ$ were combined with
data obtained by Mathewson, Ford \& Buckhorn (1992); the combined sample
was severely trimmed in order to obtain a homogeneous all--sky sample of 
1289 {\it field} objects, extending to $cz \simeq 6500$ \kms. This sample 
(SFI) is complemented by several hundred additional objects, which extends 
to higher redshifts with lesser degree of completeness. This extended sample, 
which has no significant sky coverage bias but rapidly decreasing completeness 
to $cz \sim 9500$ \kms, will be referred to as `SFI$+$'. These two field 
samples are completely independent on the sample of cluster galaxies (SCI), 
presented in Giovanelli \etal (1997a,b). A study of the SCI dipoles is 
presented in a complementary paper (Giovanelli \etal 1998).

The SFI and SFI$+$ samples are dense enough to allow estimates of dipoles 
for separate volume shells, centered on the LG. However, the windowing of 
such shells can introduce bias in the results. Such bias can be avoided 
if (a) the selection is done by observed redshift $cz$, when the {\it 
inverse} TF relation is used to estimate peculiar velocities, or if (b) 
the selection is done by TF distance $cz_{tf} = cz - V_{pec}$, when the 
{\it direct} TF relation is used (for details see Freudling \etal 1995). 
In case (b), however, it is necessary to correct the derived peculiar 
velocities for the so--called ``inhomogeneous Malmquist bias'' (IMB). 
We have computed the IMB at the location of each sample galaxy by estimating 
locally the gradient of the density field, as obtained from a redshift 
catalog (see Freudling \etal 1994 for details). Such a catalog yields 
IMB corrections of quality that decreases rapidly with distance. For the 
SFI$+$ sample, which is deeper than the strict SFI, IMB's are not available, 
and the dipole calculations are carried out only with peculiar velocities
obtained using the inverse TF approach.

\section {Dipole Results}  

Since we are 
interested in the comparison with ${\bf V}_{cmb}$, ${\bf V}_{lp}$ and
other dipole determinations, we shall estimate the dipole of the {\it reflex} 
motion of the LG with respect to our cluster set.
If $-V_i$ is the peculiar velocity of the $i$--th galaxy in the sample,
and $\epsilon_i$ is the uncertainty on that quantity, we solve for the
vector ${\bf V}_d$ of the dipole moment by minimizing the merit function
\be 
\chi^2 = \sum_i w_i
\Big({V_i - {\bf V}_d \cdot {\bf \hat r}_i\over \epsilon_i}\Big)^2  
\ee
where $\hat {\bf r}_i$ is the unit vector in the direction of the $i$--th 
galaxy and $w_i$ is a weight. $\epsilon_i$ is obtained from the TF scatter
function we have obtained for the well--determined TF template relation of
cluster galaxies (Giovanelli \etal 1997b), which varies with velocity
width $W$ as: $-0.325 (\log W - 2.5) + 0.32$ mag. Our adoption is justified 
because the $\epsilon_i$ obtained for cluster
galaxies is not affected by the cluster motions, which had been corrected
for before the TF scatter function was estimated. Such correction is much more
difficult for field galaxies, as it would require precise knowledge, point by 
point, of the unsmoothed peculiar velocity field. As shown by Giovanelli 
\etal (1997b), there is no dependence of the scatter amplitude on the 
distance of the galaxy from cluster centers: we thus feel justified in
adopting the cluster galaxy scatter function for the field galaxies.

The weights $w_i$ are intended to provide a correction that accounts for the
fading selection function of the sample with increasing distance. For
computations of dipoles of galaxies within shells, the application of a
weight $w_i$ is of limited impact. However, in the calculation of the dipole 
and bulk flow of the field over volumes including a large range in distances,
the application of weights is necessary in order to obtain an estimate 
which is independent on the particular selection function of the sample. 
The global motion of a volume bound by a top hat window can be approximated 
by using weights which are proportional to $r^3_n$, where $r_n$ is the 
distance to the $n$--th nearest neighbor to galaxy $i$ in the sample; 
$n$ is a number usually chosen between 3 and 9. We have computed
such weights for $n=4$, which matches the estimated accuracy of the approach 
with computational ease.

Table 1 lists the amplitude and apex galactic coordinates of SFI dipoles,
computed for a variety of cases and subsamples. The calculations are carried
out with $V_{pec}$'s obtained using the inverse TF relation (solutions 1--13)
and with IMB--corrected $V_{pec}$'s obtained using the direct TF relation
(solutions 14--19). The latter are only computed for the SFI sample for the
reason given in Section 2. The former are computed for both the SFI 
(odd--numbered solutions 1--11) and for the SFI$+$ samples (even--numbered 
solutions 2--12 and 13), as indicated in col. 1. Uncertainties on the dipole 
parameters are not estimated from the formal errors of the fit, but rather 
from a replacement bootstrap procedure, whereby 500 synthetic data sets are 
used, each with 63\% of the objects randomly chosen among those in the 
original data set and the remaining 37\% of the entries being duplicated 
ones. Dipole amplitudes are corrected by the ``error bias'' discussed by Lauer
\& Postman,
which in our case is of moderate or negligible importance on the results.
Each line of Table 1 gives the dipole solution for a subset of the data,
windowed in $cz$ (for the inverse TF) or in $cz_{tf}$ (for the direct TF)
as indicated. For the global solutions (1, 2 and 14) we give separately
the dipole parameters estimated for $[w_i]\equiv 1$ (solutions labelled 
`a') and for equal volume weights as described above (solutions labelled
`b'); the other solutions are averages of the two cases, albeit the
differences between the two averaged values are usually quite small and
well within the uncertainties of each determination. The number of galaxies 
used in each solution (listed in column 2) differs between the direct and 
inverse solutions for a given shell, as the windowing is carried out for 
different variables ($cz_{tf}$ and $cz$, respectively). A small fraction
of objects ($\sim$ 1--2\% ) with large ($> 1.5$ mag), possibly spurious 
magnitude offsets from the adopted TF relations were excluded in the 
calculations of dipole parameters.

\begin{deluxetable}{lrrr}
\tablewidth{0pt}
\tablenum{1}
\tablecaption{SFI Dipole Solutions}
\tablehead{
\colhead{Set} & \colhead{N$_c$} & \colhead{ V}  & \colhead{ $(l,b)$}      \nl
\tablevspace{5pt}
 &  & \colhead{km s$^{-1}$} & \colhead{$^\circ$}  
}
\startdata
{\bf Inverse TF:}                                &      &             &                  \nl
1a. 0--6500  				        & 1112 & 349$\pm$052 & (258,+40)$\pm$11 \nl
1b. 	  				        & 1112 & 433$\pm$082 & (269,+24)$\pm$19 \nl
2a. 0--6500 {\bf +}        			& 1631 & 391$\pm$039 & (255,+30)$\pm$08 \nl
2b. 	  				        & 1631 & 454$\pm$091 & (259,+25)$\pm$17 \nl
3.  0--2000	  			        &  163 & 215$\pm$096 & (268,+64)$\pm$26 \nl
4.  0--2000 {\bf +}  			        &  275 & 270$\pm$080 & (245,+49)$\pm$19 \nl
5.  1500--3500  			        &  379 & 370$\pm$071 & (265,+23)$\pm$14 \nl
6.  1500--3500 {\bf +}  		        &  549 & 410$\pm$069 & (255,+21)$\pm$12 \nl
7.  2500--4500  			        &  417 & 670$\pm$072 & (256,+18)$\pm$12 \nl
8.  2500--4500 {\bf +}			        &  580 & 620$\pm$076 & (255,+15)$\pm$11 \nl
9.  3500--5500  			        &  499 & 603$\pm$083 & (263,+24)$\pm$13 \nl
10. 3500--5500 {\bf +}			        &  689 & 585$\pm$092 & (265,+19)$\pm$13 \nl
11. 4500--6500  			        &  435 & 450$\pm$099 & (269,+21)$\pm$16 \nl
12. 4500--6500 {\bf +}			        &  635 & 544$\pm$098 & (270,+16)$\pm$15 \nl
13b. 5500--9500 {\bf +}	 		        &  506 & 620$\pm$128 & (274,+21)$\pm$19 \nl
{\bf Direct TF, IMBC:}                          &      &             &                  \nl
14a. 0--6500  				        & 1139 & 430$\pm$047 & (260,+38)$\pm$09 \nl
14b. 	  				        & 1139 & 430$\pm$079 & (262,+31)$\pm$14 \nl
15. 0--2000  				        &  113 & 258$\pm$084 & (268,+57)$\pm$29 \nl
16. 1500--3500  			        &  336 & 422$\pm$078 & (272,+25)$\pm$18 \nl
17. 2500--4500  			        &  398 & 483$\pm$091 & (269,+20)$\pm$14 \nl
18. 3500--5500  			        &  525 & 597$\pm$082 & (259,+28)$\pm$12 \nl
19. 4500--6500  			        &  552 & 609$\pm$070 & (252,+28)$\pm$13 \nl

\enddata
\end{deluxetable}

The parameters of the dipole solutions 3--13 and 15--19, as listed in Table 1, 
are displayed in Figure 1. The amplitudes are shown in panel 1(a): inverse 
TF solutions are identified by circles,
respectively unfilled and filled for the SFI and SFI$+$ samples, while
direct TF solutions are displayed as starred symbols. The horizontal dashed
line is the 611 \kms amplitude of the CMB dipole. Panel 1(b) shows the 
apices of the dipole solutions, plotted in galactic coordinates. The large,
crossed circle identifies the CMB dipole and the large square is the apex
of the LG motion with respect to the Abell cluster sample reported by Lauer
\& Postman.

The reflex motion of the LG with respect to field galaxies within 2000 \kms
exhibits a relatively small amplitude and appears directed towards high galactic
latitude. This is in agreement with the expectation that such motion is largely
affected by the presence of the density enhancement represented by the Local
Supercluster, centered on the Virgo cluster (M87 is at $l=284^\circ$, 
$b=+74^\circ$). As the radius of the shell increases, however, the LG
reflex motion asymptotically approaches ${\bf V}_{cmb}$, both in amplitude
and apex direction. Within the uncertainty of the measurement, the two
quantities become indistinguishable at distances larger than $\sim 4000$
\kms. This result is consistent with the determination obtained with
a completely independent cluster data set by Giovanelli \etal (1998), and
excludes with a high degree of confidence ($>99.99$\%) the possibility that 
the LG may exhibit a dipole such as reported by Lauer \& Postman, with 
respect to the contents of any shell within a distance of 8000 \kms.

\begin{figure}
\plotone{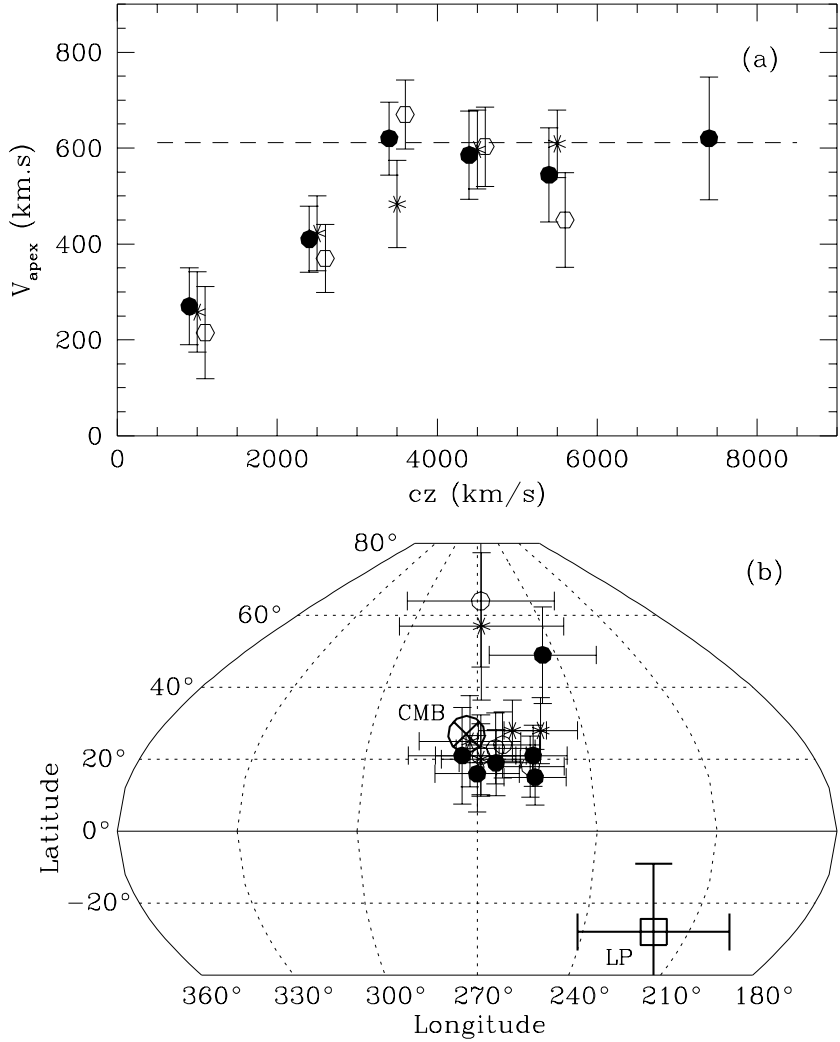}
\caption{Dipole parameters of solutions 3,5,7,9 and 11 are plotted as open
circles; those of solutions 4,6,8,10,12 and 13b are plotted as filled
circles; and those of solutions 15--19 are plotted as starred symbols.
The dashed line in panel (a) corresponds to the amplitude of the CMB
dipole, 611 \kms. The apices of the LG motion with respect to the CMB and 
the Lauer \& Postman cluster sample are labelled as `CMB' and `LP' in panel (b).} 
\end{figure}

The dipoles of the global samples (1, 2 and 14) depart from the CMB dipole
at a significant level. The equal--volume--weighted solutions (labelled `b')
are noisier than those obtained with $[w_i] \equiv 1$, an expected result
as the former give higher weight to more distant objects and errors in the 
peculiar velocity rise linearly with distance.
The difference $({\bf V}_{cmb} - {\bf V}_d)$ for any of the solutions in
Table 1 yields the bulk flow motion of the corresponding sample with respect 
to the CMB. Because many of the dipole solutions match 
so closely the CMB dipole, resulting bulk flows are quite modest, and
their directions largely unconstrained. Bulk flows associated with 
solutions 1b, 2b and 14b give an estimate of the motion, filtered by
a top hat function, of the local universe within 6500 \kms. The average
of those three solutions is 200$\pm65$ \kms towards 
$(l,b)=(295^\circ,+25^\circ)\pm 20$. This is in general agreement with
the direction of bulk flows reported in other studies (da Costa \etal 1996;
Courteau \etal 1993; Dekel 1994), but it is smaller than other determinations,
which range between 270 and 400 \kms. It agrees well in amplitude and
direction with the bulk motion with respect to clusters of galaxies
within 9000 \kms and with measured TF distances (Giovanelli \etal 1998).
It should be pointed out that the bulk flows associated with solutions
1a, 2a and 14a are somewhat larger, approaching 300 \kms amplitude in the case 
of solution 1a. These solutions do however weigh heavily nearby galaxies
and the bulk flow solutions are representative of a significantly smaller
effective volume than those for cases 1b, 2b and 14b.

In summary, we obtain that the reflex peculiar motion of the LG with respect
to field spiral galaxies approaches convergence with the CMB dipole within
6500 \kms. The dipole moment of the LG motion with respect to the outer shells 
of that volume agrees with the CMB dipole within the uncertainties. The
motion of the LG with respect to spirals within 2000 \kms is consistent with
it being influenced by the mass excess represented by the Local Supercluster.
It can be excluded to a high degree of confidence that the LG motion may exhibit 
a dipole as that reported by Lauer \& Postman, with respect to the contents of any shell 
within a distance of 8000 \kms. Finally, the bulk flow with respect to the 
CMB reference frame of a sphere of 6500 \kms radius, bound by a top hat window,
is $200\pm65$ \kms, directed towards $(l,b)=(295^\circ,+25^\circ)\pm 20$.

\acknowledgements

The results presented in this paper are based on observations carried out at
the Arecibo Observatory, which is part of the National Astronomy and 
Ionosphere Center (NAIC), at Green Bank, which is part of the National Radio 
Astronomy Observatory (NRAO), at the Kitt Peak National Observatory (KPNO), the 
Cerro Tololo Interamerican Observatory (CTIO), the Palomar Observatory (PO), 
the Observatory of Paris at Nan\c cay and the Michigan--Dartmouth--MIT 
Observatory (MDM). NAIC is operated by Cornell University, NRAO  by  
Associated Universities, Inc., KPNO and CTIO by Associated Universities 
for Research in Astronomy, all under cooperative agreements with the National 
Science Foundation. The MDM Observatory is jointly operated by the University 
of Michigan, Dartmouth College and the Massachusetts Institute of Technology 
on Kitt Peak mountain, Arizona. The Hale telescope at the PO is operated by 
the California Institute of Technology under a cooperative agreement with 
Cornell University and the Jet Propulsion Laboratory.  
This research was supported by NSF grants AST94--20505 and AST96--17069 to RG, 
AST95-28860 to MH and AST93--47714 to GW.

\newpage

\vfill
\end{document}